\documentclass[12pt]{article}
\usepackage{graphicx}
\usepackage{subcaption}
\usepackage{amsmath}
\usepackage{xcolor}
\usepackage{accents}
\usepackage{multirow}

\newcommand\pubnumber{NuPhys2016-Duffy}
\newcommand\pubdate{April 20, 2017}

\def\Title#1{\begin{center} {\Large #1 } \end{center}}
\def\Author#1{\begin{center}{ \sc #1} \end{center}}
\def\Address#1{\begin{center}{ \it #1} \end{center}}

\newcommand\pubblock{\rightline{\begin{tabular}{l} \pubnumber\\
         \pubdate  \end{tabular}}}
\newenvironment{Abstract}{\begin{quotation}  }{\end{quotation}}
\newenvironment{Presented}{\begin{quotation} \begin{center} 
             PRESENTED AT\end{center}\bigskip 
      \begin{center}\begin{large}}{\end{large}\end{center} \end{quotation}}

\def\beq{\begin{equation}}
\def\eeq#1{\label{#1}\end{equation}}
\def\eeqn{\end{equation}}

\def\beqa{\begin{eqnarray}}
\def\eeqa#1{\label{#1}\end{eqnarray}}
\def\eeqan{\end{eqnarray}}

\let\bar=\overbar

\def\Dslash{\not{\hbox{\kern-4pt $D$}}}
\def\dslash{\not{\hbox{\kern-2pt $\del$}}}

\def\msb{{\bar{\ssstyle M \kern -1pt S}}}

\newcommand{\ormu}{$1R_{\mu}$}
\newcommand{\ore}{$1R_{e}$}
\newcommand{\sth}[1]{$\sin^2\theta_{#1}$}
\newcommand{\dm}[1]{$\Delta m^2_{#1}$}
\newcommand{\dcp}{$\delta_{CP}$}
\newcommand{\numu}{$\nu_{\mu}$}
\newcommand{\nue}{$\nu_e$}
\newcommand{\numub}{$\bar{\nu}_{\mu}$}
\newcommand{\nueb}{$\bar{\nu}_e$}
\newcommand{\nub}{$\bar{\nu}$}

\newcommand\brabar{\raisebox{-4.0pt}{\scalebox{.25}{
\textbf{(}}}\raisebox{-4.0pt}{{\_}}\raisebox{-4.0pt}{\scalebox{.25}{\textbf{)
}}}}
\newcommand{\bbar}[1]{\accentset{\brabar}{#1}}
\newcommand{\bplus}[1]{\accentset{(+)}{#1}}

\begin{document}
\begin{titlepage}
\pubblock

\vfill
\Title{Current Status and Future Plans of T2K}
\vfill
\Author{Kirsty Duffy\footnote{kirsty.duffy@physics.ox.ac.uk} \\ On behalf of the T2K Collaboration}
\Address{Department of Physics, University of Oxford \\ Denys Wilkinson Building, Keble Road, Oxford, OX1 3RH, United Kingdom}
\vfill
\begin{Abstract}
T2K is a long-baseline neutrino oscillation experiment, in which a muon neutrino beam is directed over a 295 km baseline from the J-PARC facility to the Super-Kamiokande detector. This allows neutrino oscillation to be studied in two channels: disappearance of muon neutrinos and appearance of electron neutrinos. T2K has collected data using both a neutrino-enhanced and an antineutrino-enhanced beam, and these proceedings present the first T2K results using both neutrino and antineutrino oscillation data. Combining the two data sets gives the first ever sensitivity to neutrino-sector CP violation from T2K data alone, as well as the most precise T2K measurement of the other neutrino oscillation parameters.
\end{Abstract}
\vfill
\begin{Presented}
NuPhys2016, Prospects in Neutrino Physics \\ 
Barbican Centre, London, UK, December 12--14, 2016
\end{Presented}
\vfill
\end{titlepage}
\def\thefootnote{\fnsymbol{footnote}}
\setcounter{footnote}{0}
%


\section{The T2K experiment}

The T2K neutrino oscillation experiment~\cite{Abe:2011ks} uses a 30-GeV proton beam produced at the J-PARC facility in Tokai, on the east coast of Japan, to create a beam of predominantly muon neutrinos or antineutrinos (with around 1\% intrinsic contamination from electron neutrinos, and a small ``wrong-sign'' contamination). 
The neutrino beam is measured by two detectors located 280 m from the production point, ND280 and INGRID, before being directed over a 295-km baseline to the far detector, Super-Kamiokande (Super-K). 
T2K uses an off-axis `trick', in which one of the near detectors (ND280) and the far detector are placed 2.5$^{\circ}$ off axis with respect to the neutrino beam. 
By the time the beam reaches the far detector, a significant fraction of the neutrinos in the beam have oscillated into electron or tau neutrinos.

The on-axis near detector, INGRID, is composed of a 7+7 cross-shaped array of iron and scintillator detector modules. INGRID data is used indirectly in the T2K oscillation analysis to measure the beam stability and direction, and estimate the uncertainty in the neutrino flux prediction, before the ND280 data are fit.

ND280, the off-axis near detector, is used directly in the oscillation analysis to reduce uncertainties due to the neutrino flux and interaction cross sections. It is a complicated detector, made up of many subdetectors. The oscillation analysis relies in particular on information from the `tracker' region of ND280: two Fine-Grained Detectors (FGDs) interleaved with three Time Projection Chambers (TPCs) in a 0.2 T magnetic field. The FGDs provide scintillator and water targets for neutrino interactions (FGD1 is entirely composed of scintillator, while FGD2 contains both scintillator and water), with excellent vertexing and resolution close to the interaction point. The three TPCs measure interaction products leaving the FGDs with very good momentum resolution and particle identification capability.

The far detector, Super-Kamiokande~\cite{Fukuda:2002uc}, is a 50-kton (22.5 kton fiducial mass) water Cherenkov detector. It has no magnetic field, so cannot distinguish neutrino from antineutrino interactions. However, the detector is capable of very good lepton flavour identification from the pattern of Cherenkov light produced by a charged particle: it is estimated that the probability for a muon event to be misidentified as an electron is 0.7\%~\cite{Ashie:2005ik}. 

T2K can observe muon neutrino disappearance and electron neutrino appearance. The oscillation probabilities given by the PMNS matrix are~\cite{Abe:2015awa}:
\begin{align}
P(\bbar{\nu}_{\mu} \rightarrow \bbar{\nu}_{\mu}) &\simeq
	 1 - 4\cos^2\theta_{13}\sin^2\theta_{23}
	 [1-\cos^2\theta_{13}\sin^2\theta_{23}]\sin^2\frac{\Delta m^2_{32} L}{4E} \nonumber \\
	&+ \mbox{(solar, matter effect terms)} \nonumber
\end{align}
\begin{align}
P(\bbar{\nu}_{\mu} \rightarrow \bbar{\nu}_e) &\simeq 
	\sin^2\theta_{23} \sin^22\theta_{13} \sin^2\frac{\Delta m^2_{31} L}{4E} \nonumber \\ 
	&\bplus{-}\frac{\sin2\theta_{12} \sin2\theta_{23}}{2\sin\theta_{13}} \sin\frac{\Delta m^2_{21} L}{4E} 
	\times \sin^22\theta_{13} \sin^2\frac{\Delta m^2_{31} L}{4E} \sin\delta_{CP} \nonumber \\
	&+\mbox{(CP-even, solar, matter effect terms)} \nonumber
\end{align}
where the parentheses show the corresponding antineutrino oscillation probabilities. Because of the energy and baseline used for T2K, it is only sensitive to oscillations governed by the mass-squared splitting \dm{32}, and not to the so-called `solar terms', which are determined by \dm{21}. The T2K neutrino beam can be run in two configurations: `neutrino mode' (for a neutrino beam composed of mostly \numu), and `antineutrino mode' (for a beam composed of mostly \numub). This gives four `channels' that can be used to measure neutrino oscillation: \numu~disappearance, \nue~appearance, \numub~disappearance, and \nueb~appearance. These proceedings present the first T2K analysis to fit all four channels simultaneously, using a data set amounting to $7.482 \times 10^{20}$ protons on target (POT) in neutrino mode and $7.471 \times 10^{20}$ POT in antineutrino mode. This gives a precise measurement of the oscillation parameters \sth{23}, \sth{13}, and \dm{32}, as well as the first sensitivity to \dcp~from T2K alone. 

\section{Oscillation analysis strategy}

The oscillation analysis relies on models for the T2K neutrino flux (informed by external hadron production data~\cite{Abgrall:2015hmv} and in-situ measurements by INGRID and beam monitors), neutrino interaction cross sections (informed by external neutrino interaction data), and the ND280 and Super-K detector response. Using these models, data samples from ND280 and Super-K are fit simultaneously to produce an estimate of the oscillation parameters. T2K has three separate oscillation analyses, two of which take a slightly different approach to that presented here: the ND280 data are fit first, and the results of that fit propagated to Super-K for a separate oscillation fit. However, all three analyses show very good agreement in the oscillation results.

%
Only events in which a single Cherenkov ring is detected are included in the Super-K data selection for this analysis, and most events included in the data samples are expected to be quasielastic scattering interactions ($\nu_{\alpha} + n \rightarrow \alpha^- + p$, where $\alpha$ could be $\mu$ or $e$). The Super-K data are separated into sub-selections by the flavour of lepton presumed to have produced the Cherenkov ring (either electron-like or muon-like). This results in four Super-K data samples in total: neutrino-mode \ormu~(single-ring muon-like), neutrino-mode \ore~(single-ring electron-like), antineutrino-mode \ormu, and antineutrino-mode \ore.

The neutrino-mode ND280 data are separated into six selections. Three data samples are defined by the number of pions detected in the final state: \numu~CC 0$\pi$ (which is dominated by quasielastic scattering, the `signal' at Super-K), \numu~CC 1$\pi^+$ (dominated by resonant pion production, an interaction mode which forms significant background at Super-K), and \numu~CC Other (containing all other interactions). These three selections are then applied separately to neutrino interactions in FGD1 and FGD2. 

\begin{figure}[htb]
\centering
\begin{subfigure}[t]{0.49\textwidth}
	\includegraphics[width=0.9\textwidth]{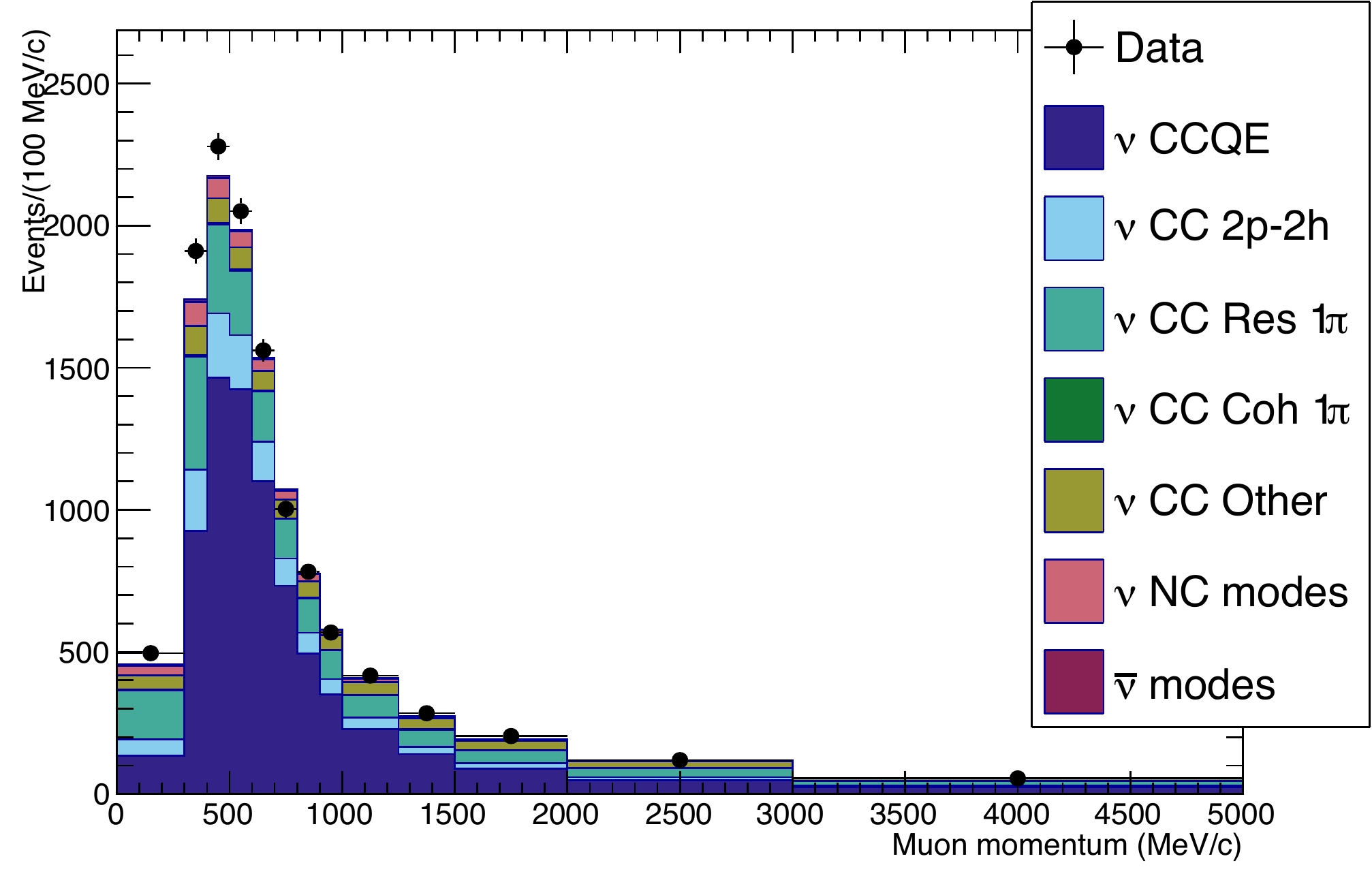}
	\captionsetup{width=0.85\textwidth}
	\caption{FGD1 \numu~CC 0$\pi$ selection ($\nu$ mode)}	
\end{subfigure}
\begin{subfigure}[t]{0.49\textwidth}
	\includegraphics[width=0.9\textwidth]{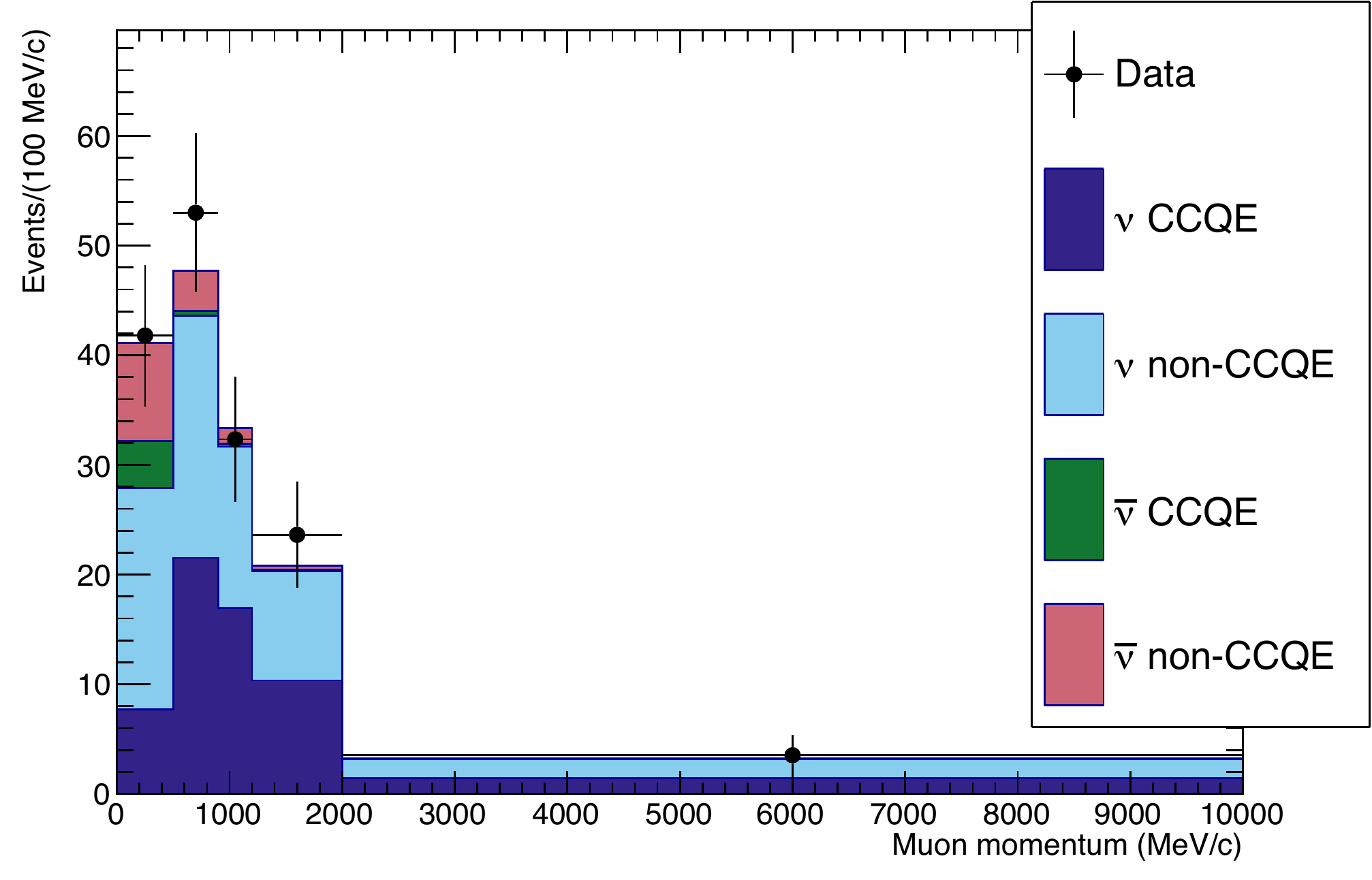}
	\captionsetup{width=0.85\textwidth}
	\caption{FGD1 \numu~CC 1-track selection (\nub~mode)}	
\end{subfigure}
\begin{subfigure}[t]{0.49\textwidth}
	\includegraphics[width=0.9\textwidth]{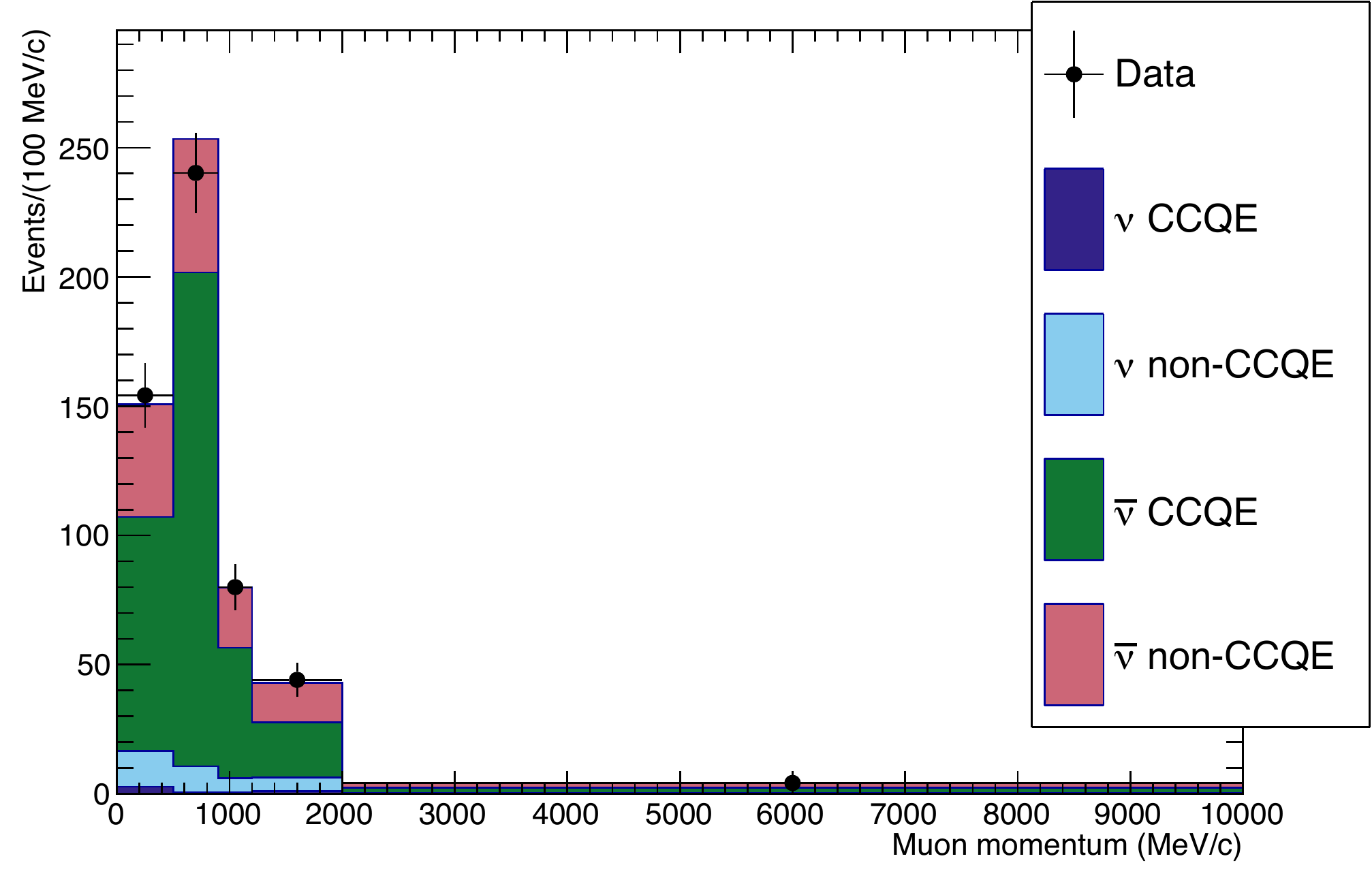}
	\captionsetup{width=0.85\textwidth}
	\caption{FGD1 \numub~CC 1-track selection (\nub~mode)}	
\end{subfigure}
\caption{A subset of the data selections used at ND280. The pre-fit prediction as a function of reconstructed muon momentum is shown as a stacked histogram, with different colours representing different neutrino interaction modes, and data points are overlaid in black.}
\label{fig:nd_data}
\end{figure}

In antineutrino mode, both \numu~and \numub~candidate interactions are selected at ND280, since there is a large wrong-sign contamination from \numu~in the antineutrino beam. Because the statistics are lower in these samples, only two categories of sub-sample are defined: \numub~(or \numu) CC 1-track (dominated by quasielastic scattering), and \numub~(or \numu) CC N-track (where N$>1$, containing mostly non-quasielastic interactions). Again, these selections are applied to interactions in FGD1 and FGD2 separately, resulting in eight antineutrino-mode data samples.
The FGD1 \numu~CC 0$\pi$, \numu~CC 1-track, and \numub~CC 1-track data and pre-fit predictions are shown in Figure~\ref{fig:nd_data}.

Including ND280 data in the fit significantly reduces the systematic uncertainty in the predicted number of events at Super-K. Measuring the `unoscillated' event rate close to the neutrino production point allows the neutrino flux and interaction cross sections to be estimated, as well as determining correlations between flux and cross-section model parameters. Overall this reduces the systematic uncertainty on the number of events in each Super-K data sample from around 12-14\% to around 5-6\%, as shown in Table~\ref{tab:syst}.

\begin{table}[t]
\begin{center}
\begin{tabular}{cc|cccc}  
\multicolumn{2}{c|}{Systematic uncertainty (\%)} & $\nu$ \ormu~& $\nu$ \ore~& \nub~ \ormu~& \nub~ \ore~\\ \hline\hline
Flux & w/o ND280 & 7.6 & 8.9 & 7.1 & 8.0 \\
Cross section & w/o ND280 & 7.7 & 7.2 & 9.3 & 10.1 \\
Flux and cross section & with ND280 & 2.9 & 4.2 & 3.4 & 4.6 \\
\multicolumn{2}{c|}{Super-K FSI/SI} & 1.5 & 2.5 & 2.1 & 2.5 \\
\multicolumn{2}{c|}{Super-K detector response} & 3.9 & 2.4 & 3.3 & 3.1 \\ 
\hline
\multirow{2}{*}{Total} & w/o ND280 & 12.0 & 11.9 & 12.5 & 13.7 \\
 & with ND280 & 5.0 & 5.4 & 5.2 & 6.2 \\
\hline
\end{tabular}
\caption{Uncertainty in the total number of events in each Super-K data sample due to different sources of systematic uncertainty.}
\label{tab:syst}
\end{center}
\end{table}

%
The data are fit using the PMNS framework for neutrino oscillation. Flat priors are used for the oscillation parameters \sth{23}, \dcp, and \dm{32}~(including a flat prior on the mass hierarchy, determined by the sign of \dm{32}), and Gaussian priors from external measurements~\cite{Agashe:2014kda} are used for the solar parameters: $\sin^22\theta_{12} = 0.846 \pm 0.021$, and \dm{21}$ = (7.53 \pm 0.18) \times 10^{-5} \mbox{eV}^2$. The data are fit twice: once with a flat prior on \sth{13}, and once with a Gaussian prior on $\sin^22\theta_{13}$ from measurements by reactor neutrino experiments~\cite{Agashe:2014kda}, $\sin^22\theta_{13} = 0.085 \pm 0.005$ (referred to as the `reactor constraint'). All nuisance parameters are removed by marginalisation (a process in which the probability distribution is integrated over nuisance parameters). 

\section{Oscillation analysis results}

\begin{figure}[htb]
\centering
\begin{subfigure}[t]{0.49\textwidth}
	\centering
	\includegraphics[width=0.8\textwidth, trim={0 16 0 0}, clip]{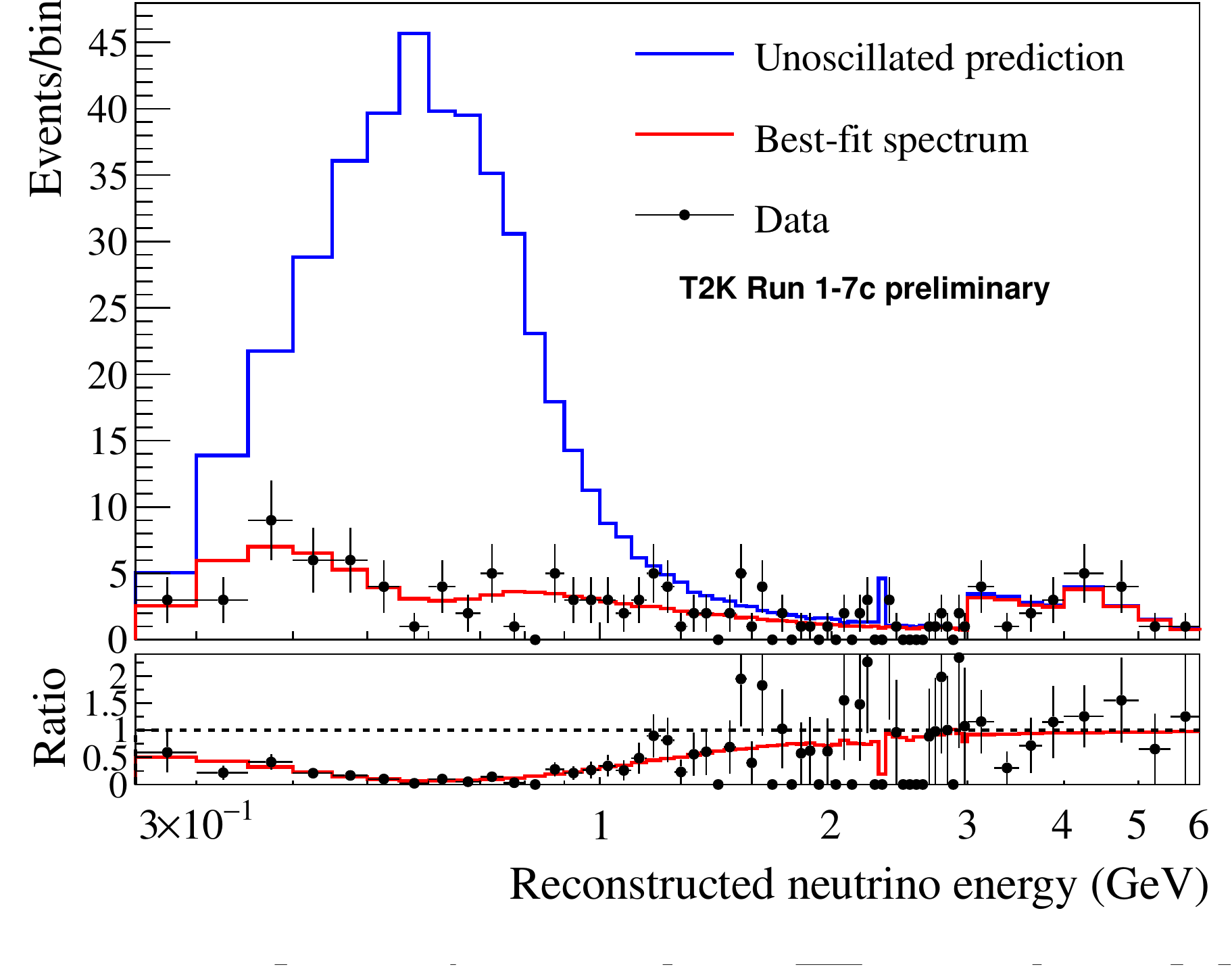}
	\caption{$\nu$-mode \ormu: 135 events}	
\end{subfigure}
\begin{subfigure}[t]{0.49\textwidth}
	\centering
	\includegraphics[width=0.8\textwidth, trim={0 16 0 0}, clip]{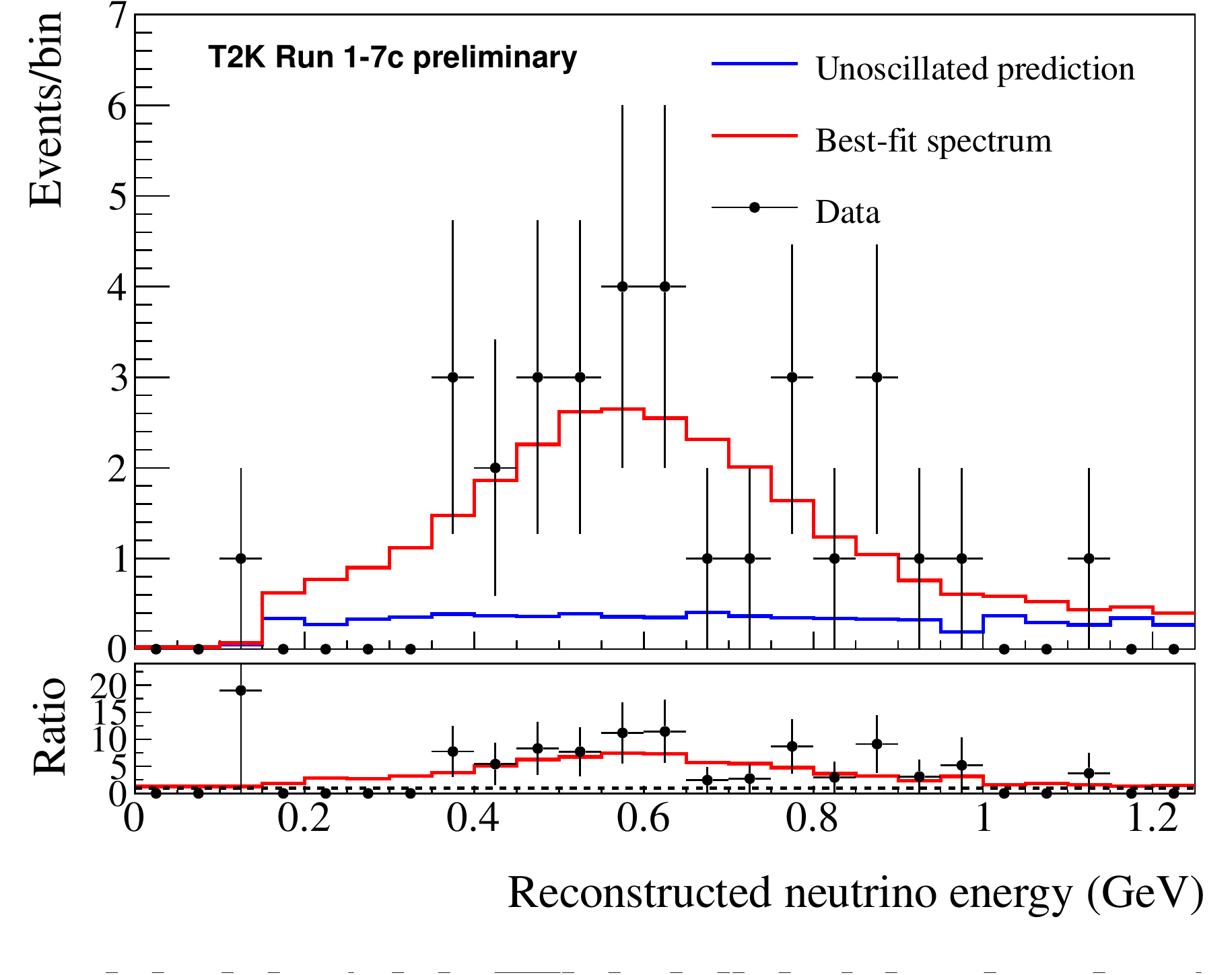}
	\caption{$\nu$-mode \ore: 32 events}	
\end{subfigure}
\begin{subfigure}[t]{0.49\textwidth}
	\centering
	\includegraphics[width=0.8\textwidth, trim={0 16 0 0}, clip]{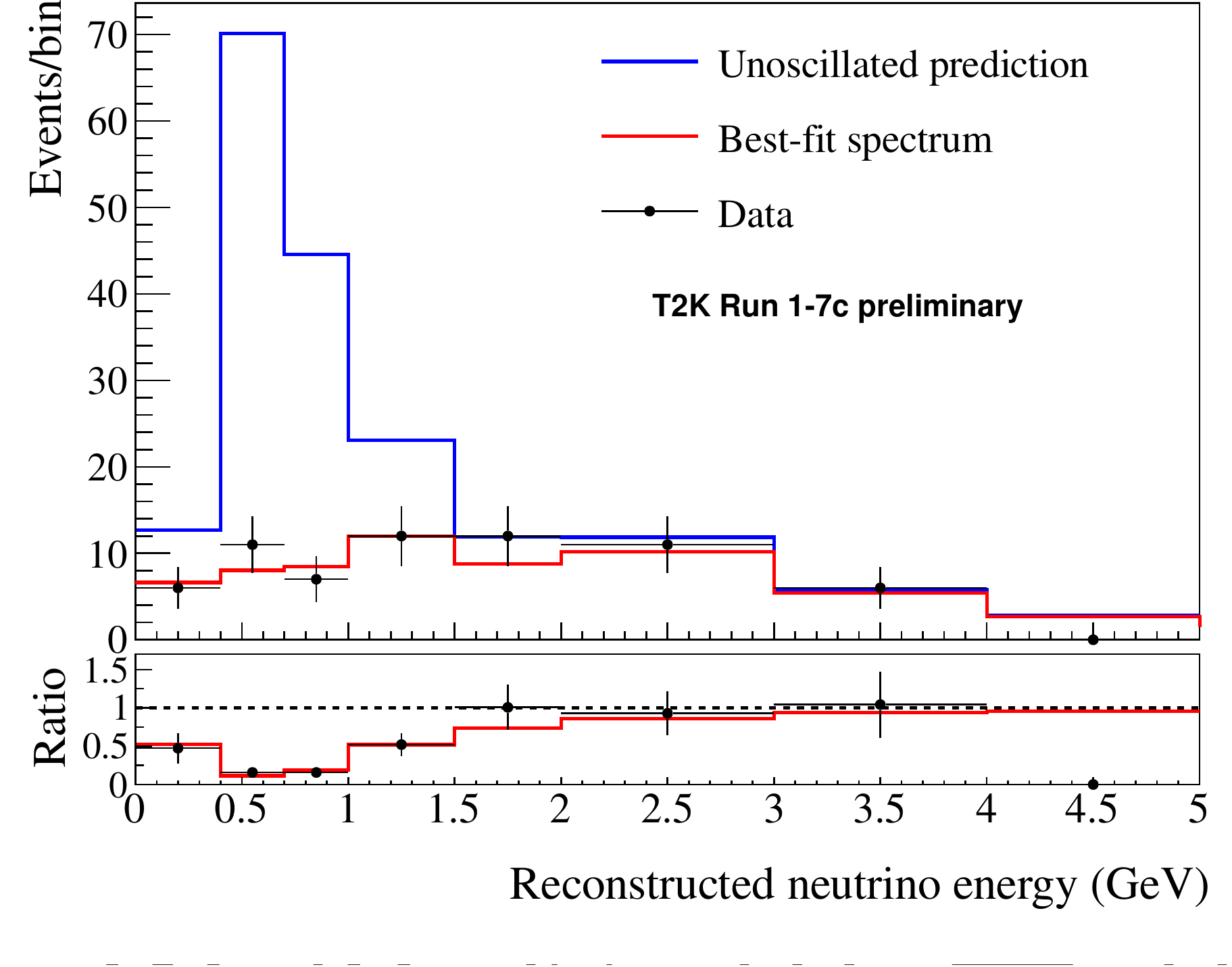}
	\caption{\nub-mode \ormu: 66 events}	
\end{subfigure}
\begin{subfigure}[t]{0.49\textwidth}
	\centering
	\includegraphics[width=0.8\textwidth, trim={0 16 0 0}, clip]{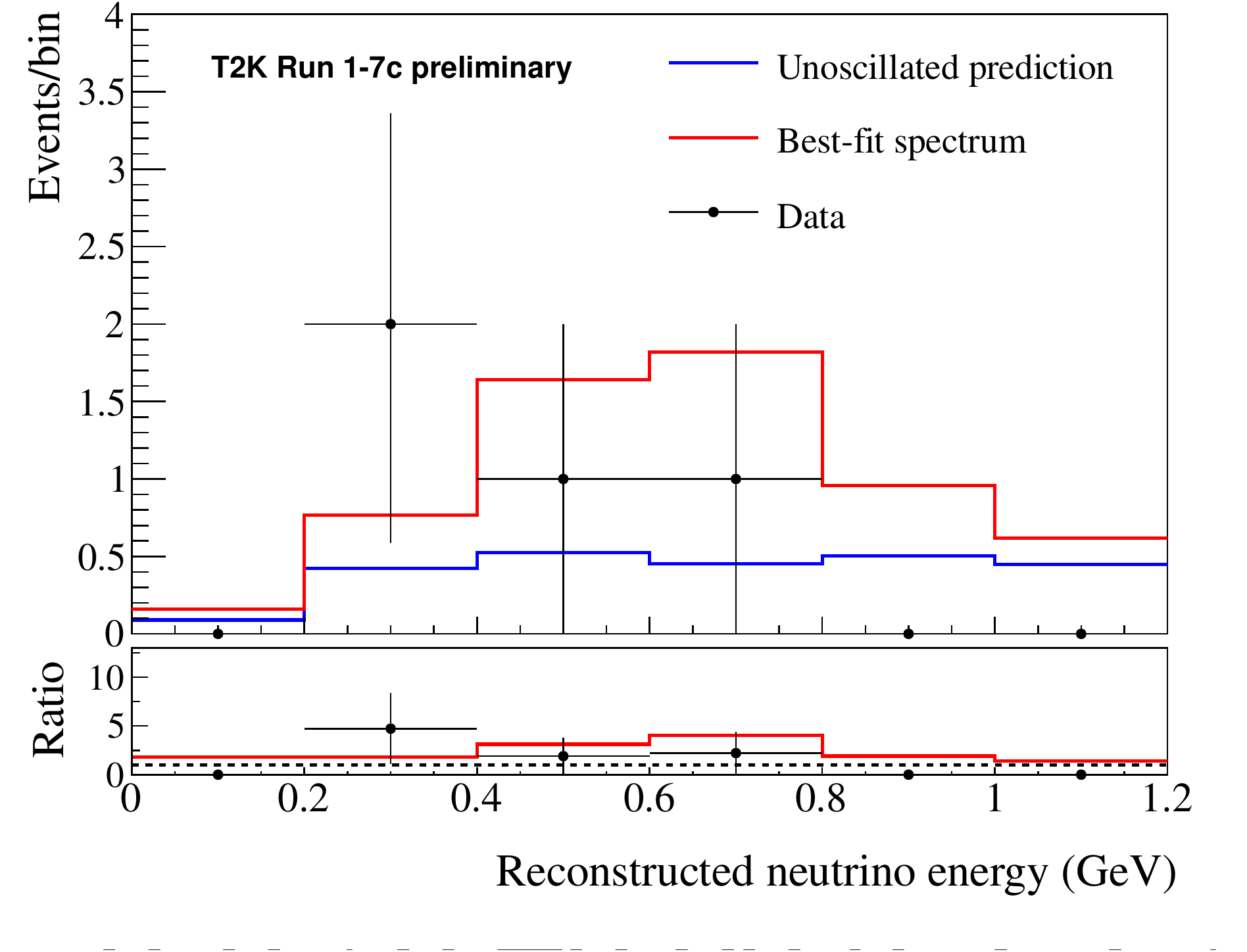}
	\caption{\nub-mode \ore: 4 events}	
\end{subfigure}
\caption{Predicted reconstructed energy spectra for the four data samples in the absence of neutrino oscillations, and after the data fit without reactor constraint on $\sin^22\theta_{13}$. The data are overlaid, and the ratio of the data and best-fit prediction to the unoscillated prediction is also shown.}
\label{fig:erec}
\end{figure}

Figure~\ref{fig:erec} shows the reconstructed neutrino energy distribution in all four Super-K data samples. The expectation in the absence of neutrino oscillation and the best-fit spectrum are also shown, as well as the ratio of data and best fit to the unoscillated expectation.

\begin{figure}[htb]
\centering
\begin{subfigure}[t]{0.49\textwidth}
	\includegraphics[width=0.97\textwidth]{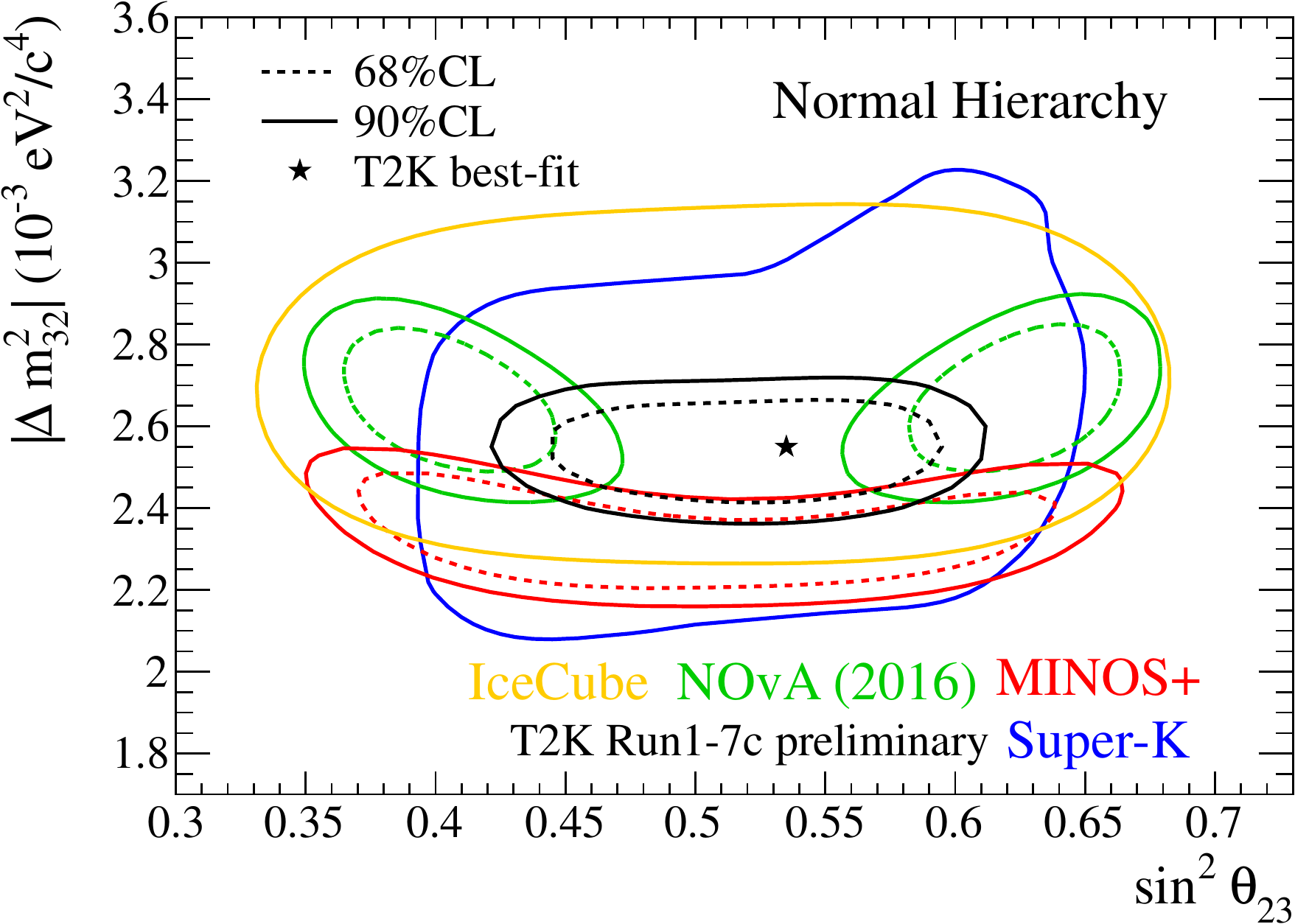}
	\caption{Normal hierarchy, \dm{32}$>$0}	
\end{subfigure}
\begin{subfigure}[t]{0.49\textwidth}
	\includegraphics[width=0.97\textwidth]{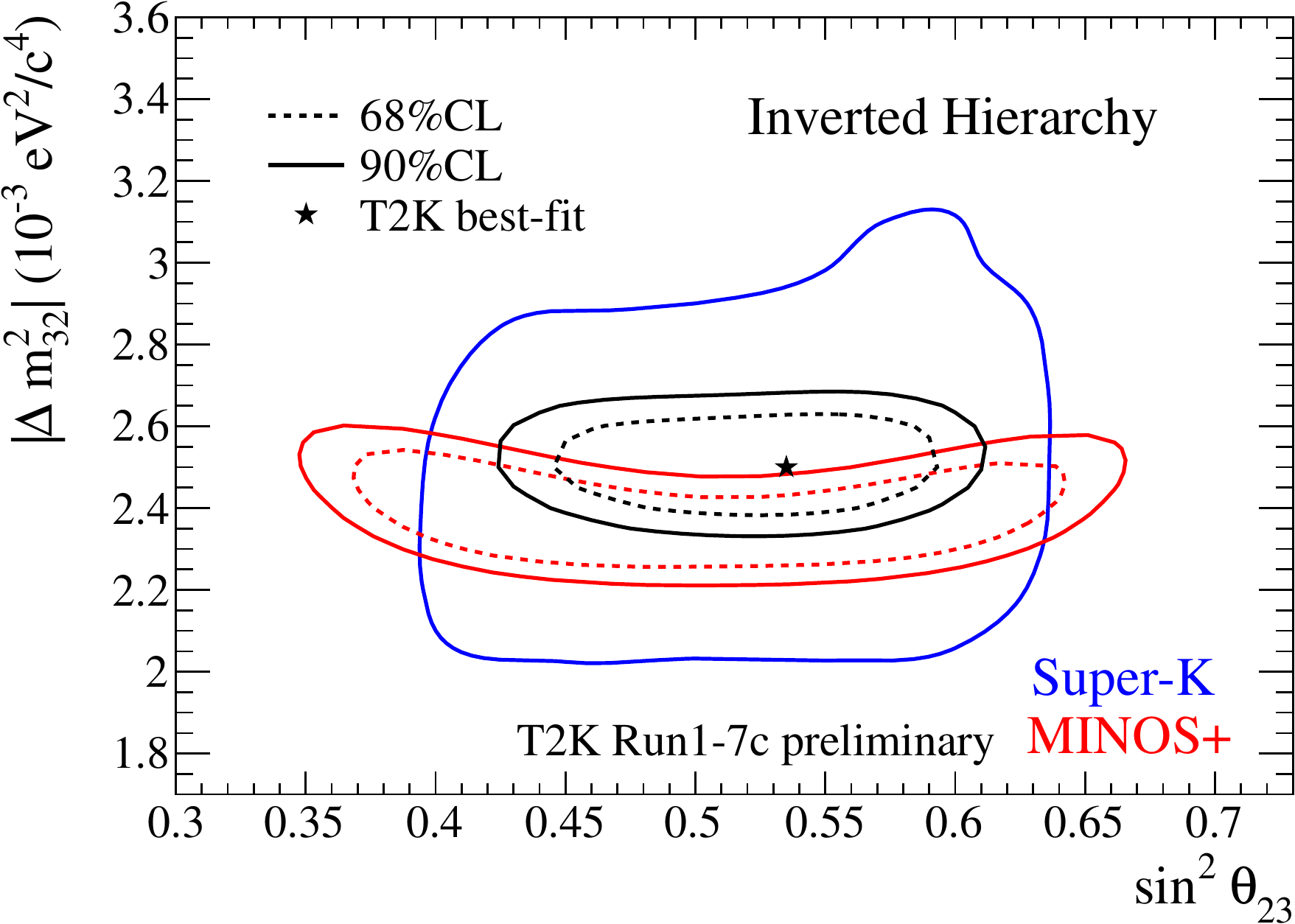}
	\caption{Inverted hierarchy, \dm{32}$<$0}	
\end{subfigure}
\caption{Constant-$(-2\Delta\ln{\mathcal{L}})$ confidence level contours in \sth{23}--\dm{32} from the T2K data fit with the reactor constraint on $\sin^22\theta_{13}$, compared to results from IceCube DeepCore~\cite{Aartsen:2014yll}, NO$\nu$A~\cite{Adamson:2017qqn}, MINOS and MINOS+~\cite{Sousa:2015bxa}, and Super-Kamiokande~\cite{Wendell:2015onk}.}
\label{fig:th23dm32_otherexp}
\end{figure}

Figure~\ref{fig:th23dm32_otherexp} shows fixed-hierarchy confidence level contours in \sth{23}--\dm{32} compared to those from other neutrino oscillation experiments. All results are consistent in both \sth{23} and \dm{32}. Results from the NO$\nu$A experiment favour non-maximal \sth{23} whereas the T2K results are consistent with maximal \sth{23}, but the both are consistent at 68\% confidence level. The MINOS and MINOS+ result prefers a slightly lower value of \dm{32}~than T2K, but -- again -- is consistent at 90\% confidence level.
Table~\ref{tab:posterior} shows the posterior probability from the T2K data fits as a function of mass hierarchy and octant of $\theta_{23}$. Both fits mildly favour the upper octant and normal hierarchy, but neither result is statistically significant.

\begin{table}[ht]
\begin{center}
\begin{tabular}{c|cc|c||cc|c} 
\multicolumn{1}{c}{} & \multicolumn{3}{c}{Without reactor constraint} & \multicolumn{3}{c}{With reactor constraint} \\ 
& \sth{23}$<$0.5 & \sth{23}$>$0.5 & Sum & \sth{23}$<$0.5 & \sth{23}$>$0.5 & Sum \\ \hline\hline 
IH (\dm{32}$<$0) & 0.16 & 0.20 & 0.36 & 0.09 & 0.19 & 0.28 \\
NH (\dm{32}$>$0) & 0.27 & 0.37 & 0.64 & 0.23 & 0.49 & 0.72 \\
\hline
Sum              & 0.43 & 0.57 & 1 & 0.32 & 0.68 & 1 \\
\hline
\end{tabular}
\caption{Posterior probability (given the T2K data and models used in the analysis) for each combination of the neutrino mass hierarchy and octant of $\theta_{23}$.}
\label{tab:posterior}
\end{center}
\end{table}

\begin{figure}[htb]
\centering
\begin{subfigure}[t]{0.49\textwidth}
	\includegraphics[width=0.97\textwidth]{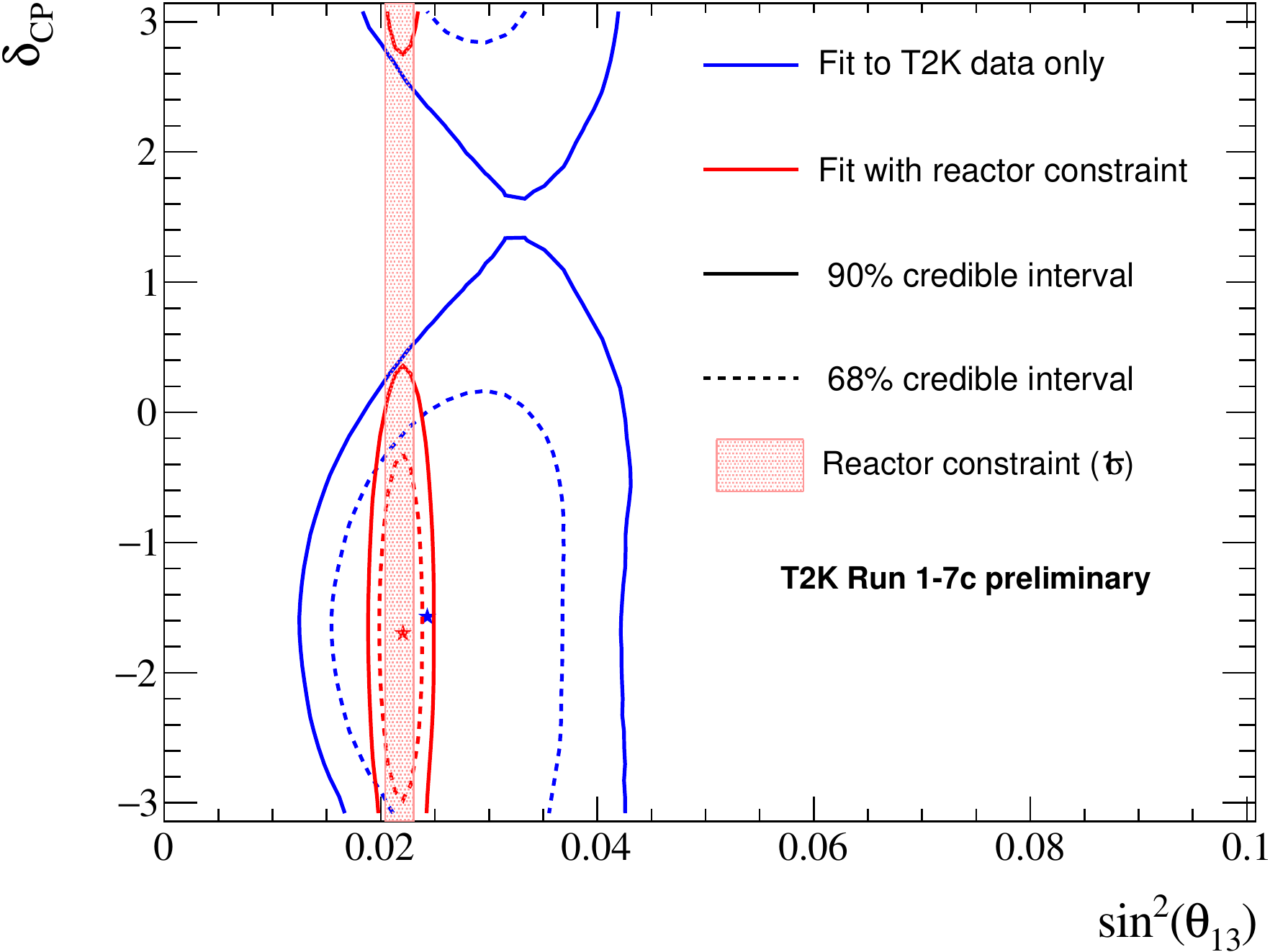}
	\captionsetup{width=0.9\textwidth}
	\caption{Credible intervals in \sth{13}--\dcp~from both fits to T2K data (with and without the reactor constraint on $\sin^22\theta_{13}$). The $\pm 1 \sigma$ band from the reactor constraint is also shown.}	
	\label{fig:th13dcp_2D}
\end{subfigure}
\begin{subfigure}[t]{0.49\textwidth}
	\includegraphics[width=0.97\textwidth]{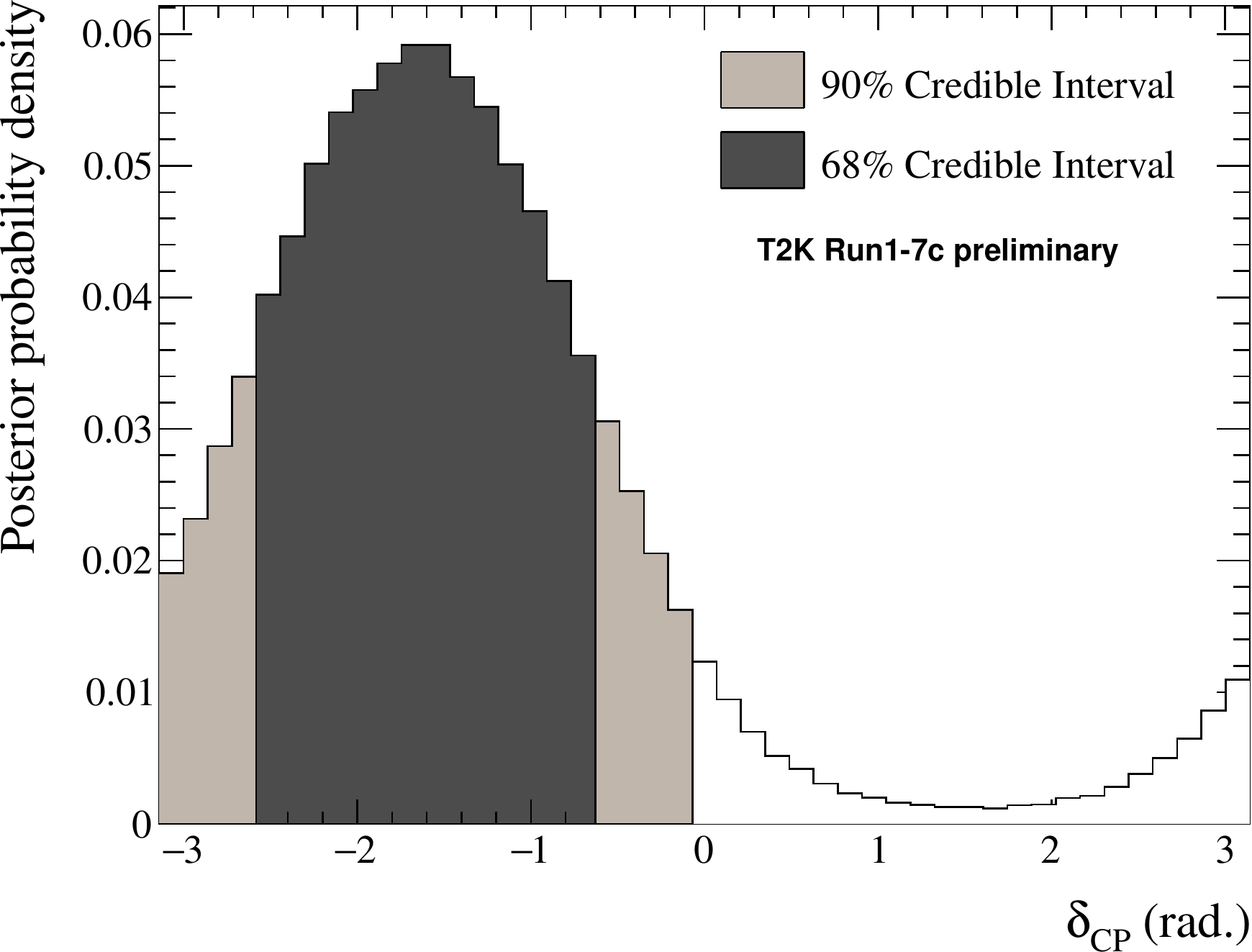}
	\captionsetup{width=0.9\textwidth}
	\caption{1D posterior probability per bin in \dcp~from the fit to T2K data with the reactor constraint on $\sin^22\theta_{13}$. The 68\% and 90\% 1D credible intervals are also shown.}	
	\label{fig:th13dcp_dcp}
\end{subfigure}
\caption{Results of the fit to T2K data in the parameters \sth{13}~and \dcp.}
\label{fig:th13dcp}
\end{figure}

The T2K measurement of \sth{13} and \dcp~is shown in Figure~\ref{fig:th13dcp}. The 2D credible intervals from data fits with and without the reactor constraint on $\sin^22\theta_{13}$ are shown in Figure~\ref{fig:th13dcp_2D}. Good agreement is seen in both fits, and the T2K measurement of \sth{13}~is consistent with the reactor measurement (shown as a red $\pm 1 \sigma$ band). Previous T2K results had no sensitivity to \dcp~when fitting without the reactor constraint, but now that antineutrino-mode data is also being included we see a 90\% closed contour.

The one-dimensional posterior probability density as a function of \dcp~from the fit with reactor constraint is shown in Figure~\ref{fig:th13dcp_dcp}. This can be interpreted as the probability -- given the T2K data and fitting model -- that the true value of \dcp~lies in a given bin on the histogram. The 68\% and 90\% 1D credible intervals are also shown. The 68\% interval contains \dcp$\in [-2.58, -0.628]$, and the 90\% interval covers \dcp$\in [-3.10, -0.07]$, both excluding the $CP$-conserving values \dcp$= 0, \pm \pi$.

This is the first time that an experimental 90\% exclusion of the $CP$-conserving values of \dcp~has been reported, but it is important to consider the potential effect of statistical fluctuations in this measurement. The sensitivity to \dcp~in T2K is driven by $\nu_e$ and \nueb~appearance. Table~\ref{tab:predevt} shows the predicted number of events in the \ore~samples for a number of different values of \dcp~and the neutrino mass hierarchy, as well as the measured number of events in each data sample. The observed number of events is most consistent with the normal mass hierarchy and \dcp$= -\pi/2$. In fact, even these parameter values underpredict the neutrino-mode \ore~sample and overpredict the antineutrino-mode \ore~sample. This implies more $CP$ violation than is physically possible in the PMNS framework, and the result is a stronger-than-expected exclusion of \dcp$=0$ and $\pm \pi$. However, this could just be due to statistical fluctuations in the two samples, which contain small numbers of events. This is important because statistical fluctuations can go both ways; if we are indeed seeing this stronger-than-expected constraint on \dcp~because of a statistical fluctuation, we may find that the \dcp~constraint gets ``worse'' as more data are collected if the fluctuation resolves or goes in the other direction.

\begin{table}[ht]
\begin{center}
\begin{tabular}{c|cc|cc} 
 & \multicolumn{2}{|c|}{$\nu$-mode \ore} & \multicolumn{2}{|c}{\nub-mode \ore} \\ \hline\hline 
Mass hierarchy & Normal & Inverted & Normal & Inverted \\
\hline
\dcp=-$\pi$/2 & 28.7 & 25.4 & 6.0 & 6.5 \\
\dcp=0 		  & 24.2 & 21.3 & 6.9 & 7.4 \\
\dcp-$\pi$/2  & 19.6 & 17.1 & 7.7 & 8.4 \\
\dcp=$\pm \pi$& 24.1 & 21.3 & 6.8 & 7.4 \\
\hline
Data          & \multicolumn{2}{|c|}{32} & \multicolumn{2}{|c}{4} \\
\hline
\end{tabular}
\caption{Number of events observed in the neutrino-mode and antineutrino-mode \ore~samples and predicted for different oscillation parameters.}
\label{tab:predevt}
\end{center}
\end{table}

\section{Summary and future prospects}

In summary, these proceedings present the first joint analysis of neutrino and antineutrino appearance and disappearance at T2K, using roughly equal amounts of protons on target in neutrino-mode and antineutrino-mode beam. This results in an extremely precise measurement of the oscillation parameters \sth{23} and \dm{32}, as well as an independent measurement of \sth{13}, all of which are in agreement with measurements by other experiments. Simultaneously analysing the oscillation of neutrinos and antineutrinos has produced the first ever experimental 90\% exclusion of the $CP$-conserving values of \dcp, \dcp=0 and $\pm \pi$, but it is important to remember that the analysis currently has low statistics. 

T2K is continuing to collect neutrino data, as well as refine the data selections and neutrino interaction models used in the oscillation analysis, in order to improve on the measurement presented here. Additionally, a number of short- and long-term analysis and detector improvements are under discussion.

One such short-term improvement is the addition of new data samples at Super-K. A new selection has been developed for single-ring electron-like events with one additional delayed Cherenkov ring due to a Michel electron from pion decay. This allows for resonant \nue~interactions which produce a $\pi^+$. The new sample is expected to add around 10\% to the statistics of the neutrino-mode \ore~sample, although the events in this sample may be less sensitive to the neutrino oscillation parameters. Updated results, including this sample, were presented at the \emph{Lake Louise Winter Institute 2017}~\cite{RajLLWI}.

Considering longer-term improvements: T2K has been approved to collect $7.8 \times 10^{21}$ POT, and is expected to do so by around 2021. A proposal is under discussion to begin T2K phase 2 in 2021 and run up until the expected start of the Hyper-Kamiokande experiment in 2026. A main ring power supply upgrade would increase the beam power, allowing T2K phase 2 to collect a predicted total of $20 \times 10^{21}$ POT. An increase in the current of the magnetic horns used to focus the beam, as well as additional Super-K samples and an expanded Super-K fiducial volume is expected to provide around an additional 50\% increase in the effective statistics. 
These statistical improvements would allow T2K phase 2 to reach 3$\sigma$ sensitivity to exclude $\sin(\delta_{CP})=0$ if \dcp=$-\pi/2$ with around $20 \times 10^{21}$ POT, assuming current systematic uncertainties. If the systematic uncertainty on the Super-K prediction can be reduced from $\sim$6\% to $\sim$4\%, then 3$\sigma$ sensitivity is expected with around $15 \times 10^{21}$ POT. As previously demonstrated, the near detector measurement is key to reducing the systematic uncertainty. Therefore, to this end a proposal is being developed to improve the systematic measurement by upgrading ND280 (in particular, by improving the acceptance of the detector).


\begin{thebibliography}{99}

\bibitem{Abe:2011ks}
  K.~Abe {\it et al.} [T2K Collaboration],
  Nucl.\ Instrum.\ Meth.\ A {\bf 659} (2011) 106
  doi:10.1016/j.nima.2011.06.067
  [arXiv:1106.1238 [physics.ins-det]].
  
\bibitem{Fukuda:2002uc}
  Y.~Fukuda {\it et al.} [Super-Kamiokande Collaboration],
  Nucl.\ Instrum.\ Meth.\ A {\bf 501} (2003) 418.
  doi:10.1016/S0168-9002(03)00425-X
  
\bibitem{Ashie:2005ik}
  Y.~Ashie {\it et al.} [Super-Kamiokande Collaboration],
  Phys.\ Rev.\ D {\bf 71} (2005) 112005
  doi:10.1103/PhysRevD.71.112005
  [hep-ex/0501064].
  
\bibitem{Abe:2015awa}
  K.~Abe {\it et al.} [T2K Collaboration],
  Phys.\ Rev.\ D {\bf 91} (2015) no.7,  072010
  doi:10.1103/PhysRevD.91.072010
  [arXiv:1502.01550 [hep-ex]].

\bibitem{Abgrall:2015hmv}
  N.~Abgrall {\it et al.} [NA61/SHINE Collaboration],
  Eur.\ Phys.\ J.\ C {\bf 76} (2016) no.2,  84
  doi:10.1140/epjc/s10052-016-3898-y
  [arXiv:1510.02703 [hep-ex]].
  
\bibitem{Agashe:2014kda}
  K.~A.~Olive {\it et al.} [Particle Data Group],
  Chin.\ Phys.\ C {\bf 38} 090001 (2014) and 2015 update.
  doi:10.1088/1674-1137/38/9/090001
  
\bibitem{Aartsen:2014yll}
  M.~G.~Aartsen {\it et al.} [IceCube Collaboration],
  Phys.\ Rev.\ D {\bf 91} (2015) no.7,  072004
  doi:10.1103/PhysRevD.91.072004
  [arXiv:1410.7227 [hep-ex]].
  
\bibitem{Adamson:2017qqn}
  P.~Adamson {\it et al.} [NOvA Collaboration],
  [arXiv:1701.05891 [hep-ex]].

\bibitem{Sousa:2015bxa}
  A.~B.~Sousa [MINOS and MINOS+ Collaborations],
  AIP Conf.\ Proc.\  {\bf 1666} (2015) 110004
  doi:10.1063/1.4915576
  [arXiv:1502.07715 [hep-ex]].

\bibitem{Wendell:2015onk}
  R.~Wendell [Super-Kamiokande Collaboration],
  PoS ICRC {\bf 2015} (2016) 1062.

\bibitem{RajLLWI} https://indico.cern.ch/event/531113/contributions/2430437/





\end{thebibliography}
\end{document}